\documentclass[10pt,letterpaper,twocolumn]{article} 

\usepackage{ol}
\usepackage{hyperref}
\usepackage{amsmath}

\newcommand{\sign}{\operatorname{sign}}
\newcommand{\arccosh}{\operatorname{arccosh}}

\begin{document}

\twocolumn[ 

\title{Rows of optical vortices from elliptically perturbing a high-order beam}

\author{Mark R Dennis}

\address{School of Mathematics, University of Southampton, Highfield, Southampton SO17 1BJ, UK}

\begin{abstract}
An optical vortex (phase singularity) with a high topological strength resides on the axis of a high-order light beam.
The breakup of this vortex under elliptic perturbation into a straight row of unit strength vortices is described.
This behavior is studied in helical Ince-Gauss beams and astigmatic, generalized Hermite-Laguerre-Gauss beams, which are perturbations of Laguerre-Gauss beams.
Approximations of these beams are derived for small perturbation, in which a neighborhood of the axis can be approximated by a polynomial in the complex plane: a Chebyshev polynomial for Ince-Gauss beams, and a Hermite polynomial for astigmatic beams. 
\end{abstract} 


] 

\noindent The light beams most often studied have cylindrical symmetry -- their intensity is invariant with rotation about the optic axis.
When such beams carry orbital angular momentum\cite{abp:oam}, such as Laguerre-Gauss (LG) \cite{absw:laguerre} and Bessel beams\cite{md:bessel}, they possess an axial optical vortex (phase singularity)\cite{nye:natural}, where the intensity vanishes, and about which the phase changes by $2\pi \ell.$ 
This integer vortex strength $\ell$ equals the beam's orbital angular momentum.
For a high-order beam ($|\ell| > 1$), the axial singularity is unstable to perturbation. 
A simple example is the addition of a small-amplitude cylindrical beam without a vortex ($\ell = 0$), whence the axial vortex unfolds into $|\ell|$ unit strength vortices which, in a transverse plane, are equally spaced on a circle centered on the axis. 
Thus rotational symmetry is broken from continuous to discrete.

Here, I want to draw attention to a natural class of perturbations, where the high-order vortex breaks up into a straight row of same sign unit strength vortices, destroying rotational symmetry.
At least three examples exhibit this: helical Ince-Gauss (IG) beams\cite{bg:igmodes,bdbg:hig}, related to gaussian modes separated in elliptic coordinates; their propagation-invariant analogue, Mathieu beams\cite{gic:mathieu}; and Hermite-Laguerre-Gauss (HLG) beams\cite{av:generalized}, generated by astigmatic transformations of LG and Hermite-Gauss (HG) beams.
IG and HLG beams are perturbations of LG beams; Mathieu beams, of Bessel beams.

The unfolding of high-order vortices into rows under certain transformations is important in the understanding of the physical structure of vortex cores\cite{roux:limitation, bd:364}.
In particular, in any synthesis of LG beams from HG modes with mode converters\cite{oc:mode}, any misalignment will lead to a row of closely spaced vortices, as will be described.

Two assumptions are made without loss of generality: the description is confined to the waist plane, and $\ell$ is positive.
A normalized LG beam, of mode order $N = \ell + 2p,$ is therefore represented\cite{absw:laguerre}
\begin{equation}
   \psi_{\mathrm{LG},\ell,p} = \frac{[2^{\ell+1}p!]^{1/2} R^{\ell} \exp(i \ell \phi)}{[\pi(\ell+p)!]^{1/2} w_0^{\ell+1}}\exp\left(\textstyle{-\frac{R^2}{w_0^2}}\right) L_p^{\ell}\left(\textstyle{\frac{2R^2}{w_0^2}}\right),
   \label{eq:lg}
\end{equation}
where $L_p^{\ell}$ is an associated Laguerre polynomial\cite{as:handbook}, $R, \phi$ are polar coordinates, and $w_0$ is the waist width.
It is proportional to $R^{\ell} \exp(i \ell \phi) = (x + i y)^{\ell}$ near the axis, like any complex high-order beam.

\begin{figure}[htb]
   \centerline{\includegraphics[width=8cm]{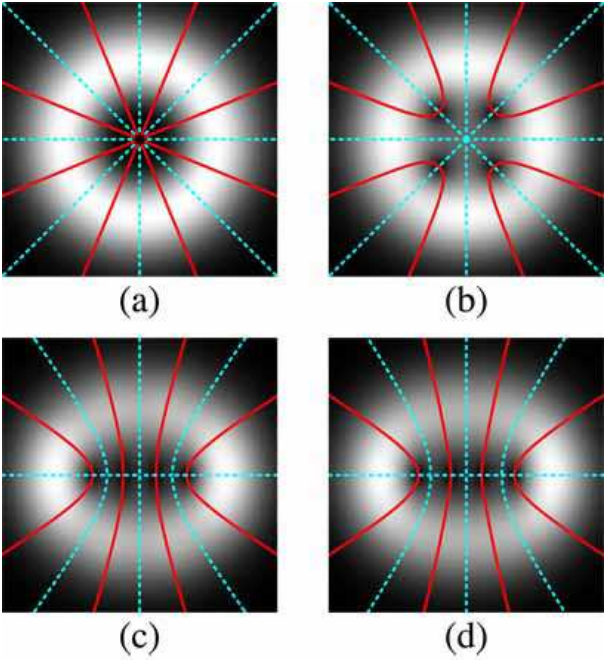}}
   \caption{   Intensity, real (solid) and imaginary (dashed) zero contours for beams in the waist plane:
   (a) LG beam $\psi_{\mathrm{LG},4,0};$ 
   (b) cylindrically perturbed $\psi_{\mathrm{LG},4,0}+0.2 \psi_{\mathrm{LG},0,0};$ 
   (c) helical IG beam $\psi_{\mathrm{IG},4,4}^2;$ 
   (d) HLG beam $\psi_{\mathrm{HLG},4,0}^{\pi/12}.$ 
   (a) is the unperturbed beam corresponding to the others. 
   In the elliptically perturbed cases (c) and (d), the order 4 vortex has unfolded into a row of four strength 1 vortices, interspersed by saddles with the same phase $i.$
   }
   \label{fig:conts}
\end{figure}

A beam's nodal structure may be understood in terms of crossings of the zero contours of its real and imaginary parts.
For $R^{\ell} \exp(i \ell \phi),$ the real and imaginary contour pattern consists of $2\ell$ alternating, regularly spaced radial lines, reflecting the $\ell$-fold rotational symmetry of the phase singularity (see Fig. \ref{fig:conts}a).
Adding a small real constant $a >0$ (approximating a vortex-free cylindrical beam) only affects the real contours (Fig. \ref{fig:conts}b), leading to vortices at positions $-a^{1/\ell} \exp(2 \pi i n /\ell)$ for $n = 1, ..., \ell.$

This contrasts with elliptically perturbed beams, where the high-order vortex breaks into a row of $\ell$ vortices of strength $+1.$
These beams are conjugation-symmetric about the $x$-axis (real (imaginary) part (anti)symmetric), with an imaginary zero at $y = 0;$ vortices occur where the real zero contours cross this line.
These vortices have the same sign, which appears to violate the sign rule\cite{fs:sign}, since they all lie on the same imaginary zero contour.
However, between each pair of real contours, another imaginary contour crosses the $x$-axis -- there is a phase saddle between adjacent vortices, which must therefore have the same sign by the extended sign rule\cite{freund:saddles}.
Examples are shown in Fig. \ref{fig:conts}c and d. 

The presence of these saddles illustrates a general feature: upon perturbation, a high-order vortex unfolds, not only to $\ell$ unit strength vortices, but also $\ell - 1$ saddle points.
This is because the Poincar{\'e} index of phase gradient (current) must be conserved; a vortex, of whatever strength, is an index $+1$ circulation, so must be balanced by $\ell - 1$ index $-1$ phase saddle points\cite{nhh:tides,freund:saddles}.
In perturbation by a constant, these saddles remain degenerate at the origin.
These observations will now be justified for helical IG beams, Mathieu beams, and HLG beams.

For the discussion of IG beams, the notation of Ref. \onlinecite{bg:igmodes} will mostly be adopted.
In the waist plane, the real and imaginary parts of helical IG beams are standing IG modes (real part symmetric, imaginary part antisymmetric).
Each mode is a product of Ince polynomials\cite{bg:igmodes,arscott:periodic} in elliptic coordinates $u, v;$ contours of constant $u$ ($v$) are confocal ellipses (hyperbolas) with foci at $y = 0, x = \pm f_0$ ($u + i v = \arccosh((x + i y)/f_0)$).
The dimensionless ellipticity parameter $e = 2f_0^2/w_0^2$ is used.
When $e = f_0 = 0,$ IG beams are LG beams (elliptic coordinates become polar), and as $f_0, e \to \infty$ they become cartesian HG beams.
A helical IG beam may therefore be treated as an elliptic perturbation of a LG beam for $f_0, e > 0.$
Real IG modes are written as a product of $\exp(-R^2/w_0^2)$ and a pair of Ince polynomials in $v$ and $iu;$ they therefore have zeros on ellipses and hyperbolas.
The helical IG beam of ellipticity $e$ corresponding to $\psi_{\mathrm{LG},\ell,p}$ is therefore\cite{bg:igmodes}
\begin{multline}
   \psi_{\mathrm{IG},N,\ell}^e =  [ \mathcal{A}^e_{C^{\ell}_{N}} C^{\ell}_{N}(i u,e) C^{\ell}_{N}(v,e)  \\
   + \mathcal{A}^e_{S^{\ell}_{N}} S^{\ell}_{N}(i u,e) S^{\ell}_{N}(v,e)] \exp(-R^2/w_0^2),
   \label{eq:hig}
\end{multline}
where the $\mathcal{A}$ denote appropriate normalization constants (found in Ref. \onlinecite{bkm:lie7} Eq. (3.14)).
The Ince polynomials $C^{\ell}_{N}, S^{\ell}_{N}$ are finite trigonometric series solving Ince's equation\cite{arscott:periodic}; $C^{\ell}_{N}$ is a sum of cosines, $S^{\ell}_{N}$ of sines.
In particular, $C^{\ell}_N(i u,e)$ is a sum of $\cosh$ terms, $S^{\ell}_N(i u,e)$ of $i \sinh$ terms, so the antisymmetric part is imaginary.

When $e >0,$ the pattern of real and imaginary zeros of a high-order vortex therefore unfolds into a system of confocal hyperbolae, as shown in Fig. \ref{fig:conts}c, given by the zeros of $C^{\ell}_{N}(v,e)$ and $S^{\ell}_{N}(v,e).$
These zeros ($\ell$ for $C,$ $\ell - 1$ for $S$) alternate, since both functions are eigenfunctions of Ince's equation, which is of Sturm-Liouville type: between adjacent zeros of $S^{\ell}_{N}(v,e),$ $C^{\ell}_{N}(v,e)$ must have a zero\cite{ch:methods1}.
Thus helical IG beams possess a straight row of alternating equal-sign zeros and saddles between their foci, with positions corresponding to the zeros of $C^{\ell}_{N}(\arccos(x/f_0),e)$ and $S^{\ell}_{N}(\arccos(x/f_0),e).$

The transition from LG to helical IG beams has been described as a perturbation; to see how the high-order vortex becomes a row, it is instructive to study the limit of small $e.$
As $e \to 0,$ Ince's equation tends to the equation of simple harmonic motion, and $C^{\ell}_{N}(v,e) \to \cos(\ell v), S^{\ell}_{N}(v,e) \to \sin(\ell v).$
Thus, for small $e = 2f_0^2/w_0^2,$
\begin{multline}
   \psi_{\mathrm{IG},N,\ell}^e \approx 
   \mathcal{A} \left( \cos(\ell v) \cosh(\ell u) + i \sin(\ell v) \sinh(\ell u)\right)  \\
   = \mathcal{A} \cosh(\ell (u+i v)) = \mathcal{A} T_{\ell}((x+iy)/f_0)),
   \label{eq:higapprox}
\end{multline}
where $T_{\ell}$ is a Chebyshev polynomial of the first kind\cite{as:handbook} (in the limit, the normalizations of $C$ and $S$ are equal\cite{bkm:lie7}).
A neighborhood of the origin, of order $f_0,$ may therefore be approximated by a polynomial in the complex plane, whose zeros, which scale with $f_0,$ are all on the real axis.
A consequence of this representation is that all zeros have sign $+1$ (they are not poles), and their cores are isotropic\cite{ss:parameterization}.
This isotropy is somewhat surprising, as one might expect the singularity phase structure to be squeezed as the row contracts; the isotropy is ensured by the complex analytic approximation\cite{dennis:thesis}.

The argument for Mathieu beams is almost identical.
The vortex (saddle) positions on the row is given by the zeros of (anti)symmetric Mathieu functions, which have the same limiting behavior as Ince polynomials.

HLG beams are the final example.
They occur in experiments when a HG or LG beam undergoes an astigmatic transformation\cite{av:generalized}, due, for instance, to a cylindrical lens or variable-phase mode converter\cite{oc:mode}, the perturbation parameter being the orientation angle $\alpha$ of the lens.
(In the language of Ref. \onlinecite{oc:mode}, $\theta = \pi/2$ and $\phi = \alpha.$)
They can be mathematically understood using the analogy between gaussian beams in the waist plane and the quantum 2D harmonic oscillator\cite{da:analogies}: HG states correspond to linear, LG to circular, and HLG to elliptic orbits.
Using Schwinger's analogy between the 2D harmonic oscillator and quantum spin\cite{bargmann:representations}, HLG beams may be written as sums of LG beams whose coefficients are spin rotation matrix elements (Wigner $d$-functions).
Therefore, the LG beam $\psi_{\mathrm{LG},\ell,p}$ is perturbed by $\alpha$ to 
\begin{equation}
   \psi_{\mathrm{HLG},\ell,p}^{\alpha} = \sum_{m=-N/2}^{N/2} d^{N/2}_{\ell/2, m}(2\alpha) (-1)^{m -|m|} \psi_{\mathrm{LG},2m,N/2-|m|}
   \label{eq:hlg}
\end{equation}
where $N = \ell+2p$ and the $d$-function $d^{j}_{m',m}$ is given in Ref. \onlinecite{sakurai:modern} Eq. (3.8.33).
Note that for $\alpha = \pi/4,$ $\psi_{\mathrm{HLG}}$ is a HG beam, and an alternative representation of Eq. (\ref{eq:hlg}) has $\alpha$ replaced by $\pi/4 - \alpha,$ $\psi_{\mathrm{LG},2m,N/2-|m|}$ by $\psi_{\mathrm{HG},N/2+m,N/2-m}$ and different phase factors.
This second form is equivalent to that given in Ref. \onlinecite{av:generalized} Eq. (8) in terms of Jacobi polynomials, and for $\alpha = \pi/4,$ in Ref. \onlinecite{da:analogies} Eq. (19) using $d$-functions.
Rows of vortices have been experimentally observed for these beams\cite{av:generalized,oc:mode}.

Unlike IG beams, HLG beams cannot be expressed in terms of a separable coordinate system (since only cartesian (HG), polar (LG) and elliptic (IG) are possible\cite{bkm:lie7}). 
Thus, vortex breakup into a row cannot be justified using Sturm-Liouville analysis.
Furthermore, for large enough $\alpha,$ the negative vortices can appear on the row, which ultimately vanishes (when $\alpha = \pi/4,$ the beam is HG).
However, the high-order vortex of $\psi_{\mathrm{LG}}$ does break into a row of vortices and saddles for small $\alpha,$ and in fact, may be approximated by a complex analytic function.

Approximation of Eq. (\ref{eq:hlg}) requires that $R,\alpha \approx 0.$
To first term in $R,$ Eq. (\ref{eq:lg}) is
\begin{equation}
   \psi_{\mathrm{LG},\ell',p'} \approx \frac{[(|\ell'|+p')!]^{1/2} (x + i \sign(\ell') y)^{|\ell'|}}{(\pi p'!)^{1/2}|\ell'|!w_0^{|\ell'|+1}},
   \label{eq:lgapprox}
\end{equation}
($\ell$ may be positive or negative, since all signs appear in Eq. (\ref{eq:hlg})).
For small $\alpha,$ $d^{j}_{m',m}(2\alpha)$ can be approximated
\begin{equation}
   d^{j}_{m',m}(2\alpha) \approx \left[\frac{(j+m')!(j-m)!}{(j-m')!(j+m)!}\right]^{1/2} \frac{(-1)^{m'-m} \alpha^{m'-m}}{(m'-m)!},
   \label{eq:dapprox}
\end{equation}
when $m' \ge 0, m' \ge m$ (when $m \ge m',$ then $m, m'$ are reversed and there is no sign factor).
Substituting these into Eq. (\ref{eq:hlg}), and using the leading-order scaling $R^2/\alpha \sim$ constant, it can be shown that
\begin{equation}
   \psi_{\mathrm{HLG},\ell, p}^{\alpha} \approx \left[\frac{(\ell + p)!}{\pi p!}\right]^{1/2} \frac{\alpha^{\ell/2}}{w_0 \ell!} H_{\ell}\left(\frac{x+iy}{2 w_0\sqrt{\alpha}}\right)
   \label{eq:hlgapprox}
\end{equation}
where $H_{\ell}$ is the $\ell^{\mathrm{th}}$ Hermite polynomial\cite{as:handbook}.
That is, for small $\alpha$ and $R,$ a HLG beam is proportional to a Hermite polynomial in the variable $(x + i y)/2w_0\sqrt{\alpha};$ this is again a complex analytic approximation, so the $\ell$ zeros have strength one and are isotropic.
The vortices lie in a row since Hermite polynomials have real zeros.

I have described the phenomenon of the breakup of a high-order phase singularity in a rotationally symmetric beam into a straight row of equal sign, unit strength singularities, and gave explicit examples for helical Ince-Gauss and generalized Hermite-Laguerre-Gauss (astigmatic) beams.
In each case, the unfolded row of $\ell$ vortices scales by the square root of the dimensionless perturbation parameter ($e$ or $\alpha$), as this tends to zero.
In this limit, the transverse neighborhood of the beam axis is approximated by a polynomial depending on the complex variable $x+iy,$ with real zeros, automatically implying a straight row of vortices with equal sign.
The intensity near the row is also proportional to the perturbation strength (given explicitly in Eq. (\ref{eq:hlgapprox})).

Helical IG and HLG beams are distinct, since Eqs. (\ref{eq:higapprox}) and (\ref{eq:hlgapprox}) involve different polynomials; each row's relative vortex spacings are different.
It is likely that elliptic perturbations can be characterized using group theory\cite{bkm:lie7}, but this is outside the scope of the present work.

In a paraxial beam, an optical vortex is a transverse solution of Laplace's equation, so it has the form $(x+iy)^{\ell}+a(x-iy)^{\ell},$ with $|a|<1$\cite{ss:parameterization}.
Cylindrical symmetry implies an axial vortex is isotropic ($a=0$).
Isotropy of the unfolded vortices implies that the perturbation, to leading order, only involves cylindrical beams of the same sign.
An anisotropic, high-order vortex thus will unfold into anisotropic vortices, and not approximable by a function of $x+iy.$
Of course, there are many other ways an order $\ell$ LG beam may be perturbed, involving beams of the same mode order.
However, it is quite difficult to construct superpositions in this way leading to vortex rows; nevertheless, the elliptic perturbations described here are physically natural.


\begin{thebibliography}{10}
\newcommand{\enquote}[1]{``#1''}
\expandafter\ifx\csname url\endcsname\relax
  \def\url#1{\texttt{#1}}\fi
\expandafter\ifx\csname urlprefix\endcsname\relax\def\urlprefix{URL }\fi
\providecommand{\eprint}[2][]{\url{#2}}

\bibitem{abp:oam}
L.~Allen, S. M. Barnett, and M. J. Padgett, eds., 
\newblock \emph{Optical Angular Momentum}
\newblock (IoPP 2003).

\bibitem{absw:laguerre}
L.~Allen, M.~Beijersbergen, R. J. C. Spreeuw, and J. P. Woerdman, 
\newblock \pra 
\textbf{45}, 8185 (1992).

\bibitem{md:bessel}
D.~McGloin and K.~Dholakia, 
\newblock Cont. Phys. 
\newblock \textbf{46}, 15 (2005).

\bibitem{nye:natural}
J. F. Nye, 
\newblock \emph{Natural focusing and fine structure of light} 
\newblock (IoPP 1999).

\bibitem{bg:igmodes}
M. A. Bandres and J. C. Guti{\'e}rrez-Vega, 
\newblock \josaa 
\newblock \textbf{21}, 873  (2004).

\bibitem{bdbg:hig}
J. B. Bently, J. A. Davis, M. A. Bandres and J. C. Guti{\'e}rrez-Vega, 
\newblock \ol 
\newblock in press.

\bibitem{gic:mathieu}
J. C. Guti{\'e}rrez-Vega, M. D. Iturbe-Castillo, and S.~Ch{\'a}vez-Cerda,
\newblock \ol 
\newblock \textbf{25}, 1493 (2000).

\bibitem{av:generalized}
E. G. Abramochkin and V. G. Volostnikov,
\newblock J. Opt. A 
\newblock \textbf{6}, S157 (2004).

\bibitem{roux:limitation}
F. S. Roux,
\newblock \oc
\newblock \textbf{223}, 31 (2003).

\bibitem{bd:364}
M. V. Berry and M. R. Dennis,
\newblock J. Opt. A 
\newblock \textbf{6}, S178 (2004).

\bibitem{oc:mode}
A. T. O'Neil and J.~Courtial
\newblock \oc
\newblock \textbf{181}, 35 (2000).

\bibitem{as:handbook}
M.~Abramowitz and I.~Stegun, eds., 
\newblock \emph{Handbook of Mathematical Functions}
\newblock (Dover, 1965).

\bibitem{fs:sign}
I.~Freund and N.~Shvartsman, 
\newblock \pra 
\newblock \textbf{50}, 5164 (1994).

\bibitem{freund:saddles}
I.~Freund, 
\newblock \pre 
\newblock \textbf{52}, 2348 (1995).

\bibitem{nhh:tides}
J. F. Nye, J. V. Hajnal, and J. H. Hannay, 
\newblock Proc. R. Soc. A 
\newblock \textbf{417}, 7 (1988).

\bibitem{arscott:periodic}
F. M. Arscott, 
\newblock \emph{Periodic Differential Equations} 
\newblock (Pergamon Press, 1964), pp145--152.

\bibitem{bkm:lie7}
C. P. Boyer, E. G. Kalnins, and W.~Miller~Jr, 
\newblock J. Math. Phys. 
\newblock \textbf{16}, 512 (1975).

\bibitem{ch:methods1}
R.~Courant and D.~Hilbert, 
\newblock \emph{Methods of Mathematical Physics}, vol.~1
\newblock (Interscience Publishers, 1953), pp 451--456.

\bibitem{ss:parameterization}
Y. Y. Schechner and J.~Shamir,
\newblock \josaa
\newblock \textbf{13}, 967 (1996).

\bibitem{dennis:thesis}
M. R. Dennis,
\newblock Ph.D. Thesis, (Bristol University, 2001), Ch. 2.

\bibitem{da:analogies}
S.~Danakas and P. K. Aravind, 
\newblock \pra 
\newblock \textbf{45}, 1973 (1992).

\bibitem{bargmann:representations}
V.~Bargmann, 
\newblock \rmp 
\newblock \textbf{34}, 829 (1962).

\bibitem{sakurai:modern}
J. J. Sakurai, 
\newblock \emph{Modern Quantum Mechanics (Revised Edition)} 
\newblock (Addison-Wesley, 1994).

\end{thebibliography}
\end{document}